\documentstyle[osa,manuscript]{revtex}

\newcommand{\ra}{\rangle}
\newcommand{\la}{\langle}

\title{Knot complexity and the probability of random knotting}

\author{Miyuki K. Shimamura}

\address{
Graduate School of Advanced Material Sciences,
University of Tokyo 
7-3-1 Hongo, Bunkyo-ku, Tokyo 113-8656, Japan}

\author{Tetsuo Deguchi}
\address{Department of Physics,  
Ochanomizu University, 2-1-1 Ohtsuka, Bunkyo-ku, Tokyo 112-8610, Japan } 

\begin{document}

\maketitle

\begin{abstract}
The probability of a random polygon (or a ring polymer) 
having a knot type $K$ 
should depend on the complexity of the knot $K$. 
Through computer simulation using knot invariants, 
we show that the knotting probability 
decreases exponentially with respect to 
knot complexity. Here we assume that  some aspects of knot complexity   
are expressed by the minimal crossing number $C$  and the 
aspect ratio $p$ of the tube length to the diameter 
of the {\it ideal knot} of $K$, which is a tubular 
representation of $K$ in its maximally inflated state. 
\end{abstract}

\section{Introduction}

Various species of knotted polymers 
have been synthesized and observed in chemistry and biology
in the last two decades. \cite{Dean,Shishido,Rybenkov,Shaw}
Once a ring polymer is formed, its topological state is unique and invariant. 
The topological constraint on the ring polymer should be  nontrivial. 
It may restrict the available degrees of freedom in the configuration space 
of the polymer, to a great extent. Consequently, it may lead to  
a large entropic reduction, which is related to 
the probability of random knotting, as wee shall see shortly.

 For a  knot $K$, we define the knotting probability $P_K(N)$ by 
the probability that the topology of a random polygon 
with $N$ nodes is given by the knot $K$. 
 If  a ring polymer is under 
 the topological constraint of the knot $K$, 
  then the decrease of the polymer entropy  is given by 
$\Delta S_K = - k_B \log P_K(N)$, where $P_K(N)$ is the 
ratio of the volume of the configuration space 
under the topological constraint to that of no topological constraint. 
The knotting probability $P_K(N)$ should also correspond to 
the probability that a ring polymer of $N$ Kuhn units 
 have the knot $K$ when it is closed randomly during its  
synthesis. The knotting probabilities have been   
 measured as the fractions of knotted species of circular DNAs.
  \cite{Rybenkov,Shaw}   

\par 
Let us now discuss how to express the complexity of knots. 
We could classify knots completely, if we might know 
all the topological properties  
that are invariant under any continuous deformation 
of the spatial configurations.   
The topology of a given polygon can be effectively detected 
  by calculating some topological invariants 
 such as the Alexander polynomial $\Delta_K(t)$
and the Vassiliev-type invariants $v_n(K)$. 
Although the invariants are practically useful 
for computer simulations \cite{Vologodskii1,DeguchiPLA}, 
it is not easy to derive any explicit topological properties or meanings 
from them.  Let us consider 
the minimal number of crossing points in the knot diagram of a knot $K$.  
We donate it by $C$, or $|K|$ for the knot $K$. 
 The minimal crossing number $C$ 
should be a measure on the complexity of knots. 
The number $C$ is useful in studying statistical or  dynamical 
properties  of knotted ring polymers \cite{Quake,Lai}. 
There is some arguments on the mean-square radius of gyration 
of knotted ring polymers with respect to  $C$ \cite{Quake}. 
However,  $C$ is rather weak as a topological invariant.  
The number of knots that have the same number $C$ 
increases rapidly: there are 166 primes knots which have  10 crossings.

\par 
Recently the concept of ideal knots 
has attracted much interest.  
\cite{KatritchNature,KatritchBook,GrosbergPRE}.
One of the  most ideal (or elegant) geometric representations 
of a knot should be given by such a closed tube with uniform diameter 
that gives the largest ratio of the diameter to the tube length.  
We call such geometric representations {\it ideal knots}, briefly.  
For a ring polymer with a knot type $K$, 
Grosberg $et.al$ \cite{GrosbergPRE} discussed  
a topological invariant $p$ of a knot $K$, which is defined by 
 the aspect ratio of the length ($L$) to the diameter ($d$) 
of such a ring polymer of knot $K$ that is maximally inflated, 
{\it i.e.}, 
an ideal knot. Here,  $p$ is given by  $L/d$. We also denote it 
by $p_K$.  The value $p_K$  should be a measure of  knot complexity.   
It may be more powerful than $C$, 
since different knots should have different values of $p$, in general.  
Katritch {\it et. al} have obtained ideal knots for 42 
different knots \cite{KatritchNature,KatritchBook}.
There is a linear relation between the average crossing numbers of ideal knots 
and their $p$ values. \cite{KatritchNature,KatritchBook}. 
The value $p$ should be also useful 
for describing  flexible DNA knots in thermal equilibrium 
\cite{KatritchNature}. 
Furthermore, the value  $p$ should be useful for  
statistical or  dynamical researches of knotted ring polymers. 
\cite{GrosbergBook,Sheng} 

\par 
The $N$-dependence of the knotting probability 
 has been studied through simulations,  
  \cite{Vologodskii1,desCloizeaux,LeBret,Michels,Koniaris,Orlandini,DeguchiJKTR,DeguchiRevE,PLA,Miyuki-JPSJ} 
  and it is found that the probability of the unknot (the trivial knot) 
decreases exponentially with respect to  $N$:
\begin{equation}
P_0(N)=C_0 \exp(-N/N_c),
\label{knot0}
\end{equation}
where $C_0$ and $N_c$ are fitting parameters.
For some nontrivial knots ($3_1,4_1,5_1,5_2$),  
knotting probabilities have been evaluated numerically 
for several different models of 
random polygons and self-avoiding polygons.
\cite{Orlandini,DeguchiJKTR,DeguchiRevE,PLA,Miyuki-JPSJ}.
Through the simulations using the Vassiliev-type invariants, it is found that 
the probability $P_K(N)$ as a function of $N$ can be expressed as  
\begin{equation}
P_K(N)=C_K \Biggl( \frac{N}{N_K} \Biggr)^{m_K} 
\exp \Biggl( -\frac{N}{N_K} \Biggr) .
\label{knot-formula}
\end{equation}
Here $C_K$, $N_K$ and $m_K$ are fitting parameters to be determined 
from the numerical results. The expressions (\ref{knot0}) and 
(\ref{knot-formula}) 
should correspond to the asymptotic expansion of renormalization 
group arguments.  
Numerically we see that  the estimates of $N_K$ should be given by 
 almost the same value for any knot $K$ \cite{DeguchiJKTR,DeguchiRevE,PLA},  
and therefore the $N_K$'s are almost equal to  $N_c$, which depends 
on the model.  We also observe that the value $m_K$ of a knot $K$ 
should be universal for the different models \cite{DeguchiRevE}. 

\par 
In this paper, we discuss how 
the knotting probability  $P_K(N)$ of a knot $K$ 
should depend on its complexity while $N$ being fixed, or 
in short,  the knot dependence of the normalization constant $C_K$.  
Evaluating the knotting probabilities of several prime knots 
for Gaussian random polygons,  we observe a rough tendency 
 that the amplitude $C_K$ decreases exponentially with respect to $p$. 
The numerical result seems to be favorable to  
Grosberg's conjecture \cite{GrosbergBook} that 
the probability $P_K(N)$ as a function of $p$ 
should be given by 
\begin{equation}
P_K(N) \sim \exp(-N/N_c -sp).
\end{equation}
Here $s$ is a constant.  At this stage,  however, we could not judge  
whether  the conjecture be valid,  
since  the data points scatter 
outside the range of statistical errors.  
 On the other hand, we  show another explicit statistical behavior     
for a version of the  knotting probability.  
Let us define the average knotting probability $P_{ave}(N,C)$ by 
\begin{equation} 
P_{ave}(N,C) = \sum_{K: \, |K|=C} P_K(N)/A_C. 
\end{equation} 
Here the sum is over such knots that  
have the same minimal crossing number $C$, and  
 $A_C$ denotes the number of prime knots which have 
the same minimal crossing number $C$.   
 For instance, we have $A_3=A_4$=1, $A_5=2$ and $A_6=3$. 
Then, we shall see from the data that 
the average knotting probability decreases 
exponentially with respect to $C$.  
Furthermore,  if we consider 
the average of the $p$'s over such knots that have the same $C$    
\begin{equation}
\la p \ra = \sum_{K: |K|=C} p_K /A_C \, , 
\end{equation}
then we  see that the average knotting probability 
also decays exponentially with respect to the average $\la p \ra$.

\section{The method of simulations}

Using the conditional probability \cite{desCloizeaux},  
we construct a large number of 
Gaussian random polygons, say $M$ polygons, of $N$ nodes 
for $N$ = 300, 500 and 1000.  
Then, the knotting probability of a knot $K$ is evaluated by $P_K(N)=M_K/M$.  
Here $M_K$ is the number of polygons with the knot $K$, and  
$M$ is given by $M=10^5$  to each  of the three numbers of $N$.  

\par 
In order to detect the knot type of a give polygon,      
we employ three knot invariants: the determinant $|\Delta_K(t=-1)|$ 
of a knot $K$, 
the Vassiliev-type invariants $v_2(K)$ and $|v_3(K)|$ 
of the second and third degrees, respectively.
We evaluate  $M_K$, after enumerating the number of the polygons 
which have the same set of the values of the three invariants for 
the knot $K$.  
Using $|v_3(K)|$, 
we do not distinguish the chirality of the knot, {\it i.e.}, 
the right-handed knots and the left-handed ones \cite{DeguchiPLA,Polyak}.
Furthermore, 
we do not consider six knots ($8_9$, $8_{10}$, $8_{11}$, $8_{18}$, $8_{20}$ 
and $8_{21}$) in any of the simulations in the paper. They 
have the same values of the three invariants 
as those of some composite knots.

\section{Results and Discussion}

The estimates of the average knotting probability $P_{ave}(N,C)$ 
are plotted in Fig. 1 against the minimal crossing number $C$, 
 up to $C=8$. 
It is clear in  Fig. 1  
that the average knotting probability $P_{ave}(N,C)$ 
decreases exponentially with respect to  $C$.  
We remark that error bars correspond to one standard deviation 
in all the four Figures in the paper. 

\par 
Let us now  discuss the knotting probability 
in terms of the $p$ values. In Fig. 2,  
 the knotting probabilities $P_K(N)$ for some prime knots 
 are plotted against $p_K$, where $N$ is kept constant.
We note that the $p$ values of the 42 knots  
are listed in Table 1 of Ref. \cite{KatritchBook},   
which are used in the paper.  
We see in Fig. 2 that there is a rough tendency that 
the knotting probability of a prime knot decrease 
exponentially with respect to the value $p$. 
 The observation should be useful. 
 However, it seems that there is no 
  clear relation between the knotting probability $P_K(N)$ 
 and the aspect ratio $p$,   
 since  the data points of larger $p$ values deviate 
 from the possible regression line, considerably.  
Here we recall that error bars correspond to 
 one standard deviation. 

\par 
Let us discuss the knotting probability for such knots 
that have the same minimal crossing number.  
For instance, there are two knots with 5 minimal crossings: $5_1$ and $5_2$.  
 For Gaussian polygons,  the knotting probability 
 of $5_2$ is always larger than that of $5_1$.   
This is consistent with the simulation 
of the cylindrical self-avoiding polygons \cite{Miyuki-JPSJ}. 
Let us consider the three prime knots with $C=6$.  
We observe that the knotting probabilities of 
 $6_1$ and $6_2$ are almost the same, 
while that of $6_3$ is always smaller than  the other two.
 For prime knots with $C=$ 7 or 8, 
the data points are so close to each other that 
it is difficult to give any definite ranking on them. 

\par 
In terms of the average value  $\la p \ra$, which is a function of $C$, 
the estimates of the average knotting probability $P_{ave}(N, C)$ 
are expressed in Fig. 3.
 We clearly see  the exponential decay of the 
 average knotting probability $P_{ave}(N, C)$ 
 with respect to $\la p \ra$.
It is similar to  Fig. 1.  
This result shows that the entropy 
of a ring polymer with knot $K$ decreases with respect to  knot
complexity expressed in terms of $\la p \ra$ or $C$. 
Here it is also suggested  that  $\la p \ra$ should be 
 approximately linear to $C$.

\par 
Let us discuss the $N$-dependence of the knotting probability 
in terms of knot complexity. 
The ratio $P_K(N)/P_{3_1}(N)$ 
of a knot $K$ against the value $p$ 
is plotted in Fig. 4 for the three numbers of $N$:  $N=300, 500, 1000$. 
Here we note that the trefoil knot ($3_1$) is dominant 
among the nontrivial prime knots 
for the three $N$'s.   
 We find again the rough tendency that  
the ratio $P_K(N)/P_{3_1}(N)$ decays 
exponentially with respect to knot complexity $p$.
Moreover,  for any knot $K$, the ratio $P_K(N)/P_{3_1}(N)$ 
is given by almost the same value 
for the three  numbers  of $N$, with respect to error bars 
as seen in Fig. 4.  
Thus,  the ratios $P_K(N)/P_{3_1}(N)$ are  independent of $N$. 

\par 
The above observation in Fig. 4 can be explained 
by using the fitting formula (\ref{knot-formula}). 
Let us assume that for a prime knot $K$, the exponent  
 $m_K$ of eq. (\ref{knot-formula}) should be  given by almost 
 the same value.   
Then, we have $P_K(N)/P_{3_1}(N) \sim C_K/C_{3_1}$, 
which is clearly independent of $N$.  
Thus, in terms of the  formula (\ref{knot-formula}), 
 the rough exponential decay of the knotting probability 
 with respect to $p$ is closely related to 
 the  knot complexity-dependence of the amplitude $C_K$.

\vskip 24pt 
\par \noindent 
{\bf Acknowledgements} 
\par 
We would like to thank  Prof. K. Ito  for  helpful discussions.

\newpage


\begin{figure}
\caption{Average knotting probability $P_{ave}(N,C)$ 
versus the minimal crossings $C$ for 29 
 prime knots with upto $C=8$. The line is given by  
$P_{ave}(N,C) = P_{ave}(N,0) \exp(- \alpha C)$ with $\alpha=1.16$. }
\end{figure}

\begin{figure}
\caption{Knotting probability $P_{K}(N)$ with $N=500$ 
versus the aspect ratio $p$ for 29 
 prime knots with upto $C=8$.} 
\end{figure}

\begin{figure}
\caption{Average knotting probability $P_{ave}(N,C)$ 
versus the average $\la p \ra$ for 29 
 prime knots with upto $C=8$. 
 The line is given by  
$P_{ave}(N,C) = P_{ave}(N,0) \exp(- \beta \, \la p \ra)$ 
with $\beta=0.30$. }  
\end{figure}

\begin{figure}
\caption{The ratio of knotting probabilities
 $P_{K}(N)/P_{3_1}(N)$ 
versus  the aspect ratio $p$ for $N=300,500,1000$. }
\end{figure}


\begin{thebibliography}{[99]}

\bibitem{Dean} 
 F.B. Dean, A. Stasiak, T. Koller and N.R. 
Cozzarelli, J. Biol. Chem. {\bf 260} (1985) 4795-4983; 
S.A. Wasserman, J.M. Duncan and N.R. Cozzarelli, 
Science {\bf 229} (1985)171-174; Science {\bf 232} (1986) 951-960. 


\bibitem{Shishido} 
K. Shishido, N. Komiyama and S. Ikawa, 
J. Mol. Biol. $\bf{195}$ (1987) 215.


\bibitem{Rybenkov} 
V.V. Rybenkov, N.R. Cozzarelli and A.V. Vologodskii,
Proc. Natl. Acad. Sci. USA {\bf 90}, 5307 (1993)
\bibitem{Shaw}  S.Y. Shaw and J.C. Wang,Science {\bf 260}, 533 (1993).


\bibitem{Vologodskii1} A.V. Vologodskii, A.V. Lukashin, M.D.
Frank-Kamenetskii, and V.V. Anshelevich,
Sov. Phys. JETP {\bf 39},1059 (1974).

\bibitem{DeguchiPLA}
T. Deguchi and K. Tsurusaki, Phys. Lett. A $\bf{174}$, 29 (1993)

\bibitem{Quake} S.R. Quake, Phys. Rev. Lett. {\bf 73}, 3317 (1994).  


\bibitem{Lai} Y.-J. Sheng and P.-Y. Lai, Phys. Rev. E {\bf 63}, 021506 (2001).


\bibitem{KatritchNature}
V. Katritch, J. Bednar, D. Michoud, R. G. Scharein, J. Dubochet, and A. Stasiak,
Nature (London) {\bf 384}, 142 (1996)

\bibitem{KatritchBook}
A. Stasiak, J. Dubochet, V. Katritch, and P. Pieranski, 
in {\it Ideal Knots} edited by A. Stasiak, V. Katritch and L. H.
Kauffman (World Scientific, Singapore, 1998) pp. 1-19. 

\bibitem{GrosbergPRE}
A. Yu. Grosberg, A. Feigel, and Y. Rabin, Phys. Rev. E {\bf 54}, 6618 (1996).


\bibitem{GrosbergBook}
A. Yu. Grosberg, in {\it Ideal Knots} edited by A. Stasiak, V. Katritch and L. H.
Kauffman (World Scientific, Singapore, 1998) pp. 129-142. 

\bibitem{Sheng}
Y.-J. Sheng, P.-Y. Lai, and H.-K. Tsao, Phys. Rev. E {\bf 58}, R1225 (1998).




\bibitem{desCloizeaux} J. des Cloizeaux and M.L. Mehta,
J. Phys. (Paris) $\bf{40}$, 665 (1979).

\bibitem{LeBret} M. Le Bret, Biopolymers $\bf{19}$, 619 (1980).

\bibitem{Michels} J.P.J. Michels and F.W. Wiegel,
Phys. Lett. A $\bf{90}$, 381 (1982).

\bibitem{Koniaris} K. Koniaris and M. Muthukumar,
Phys. Rev. Lett. $\bf{66}$, 2211 (1991).

\bibitem{Orlandini} E. Orlandini, M.C. Tesi, E.J. Janse van Rensburg and
S.G. Whittington,
J. Phys. A: Math. Gen. {\bf 31}, 5953 (1998).

\bibitem{DeguchiJKTR} T. Deguchi and K. Tsurusaki,
J. Knot Theory and Its Ramifications $\bf{3}$, 321 (1994).

\bibitem{DeguchiRevE} T. Deguchi and K. Tsurusaki,
Phys. Rev. E $\bf{55}$, 6245 (1997).

\bibitem{PLA} M.K. Shimamura and T. Deguchi,
Phys. Lett. A  {\bf 274}, 184 (2000).  
 
\bibitem{Miyuki-JPSJ} 
M.K. Shimamura and T. Deguchi, J. Phys. Soc. Jpn. {\bf 70}, 1523 (2001). 

\bibitem{Polyak} M. Polyak and O. Viro, Int. Math. Res. Not. No.11, 445 (1994).

\end{thebibliography}
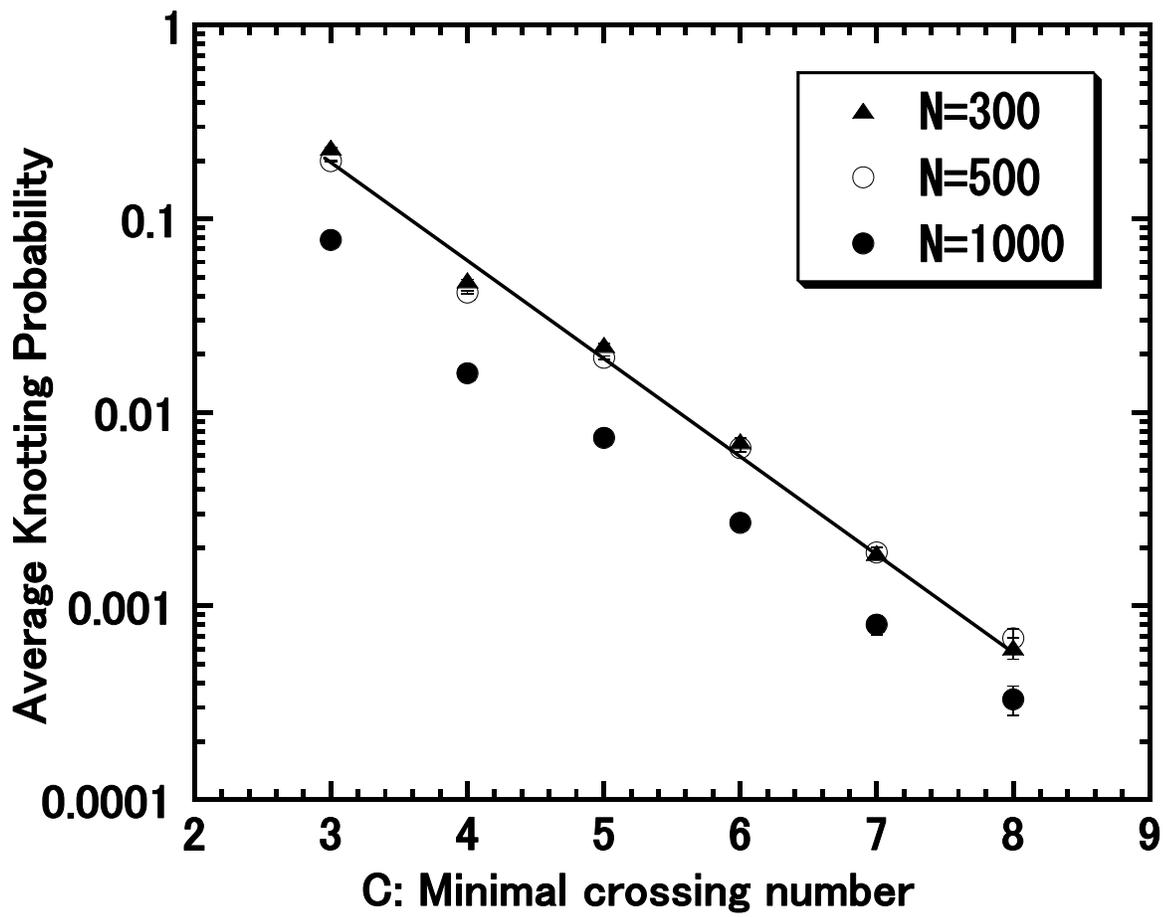
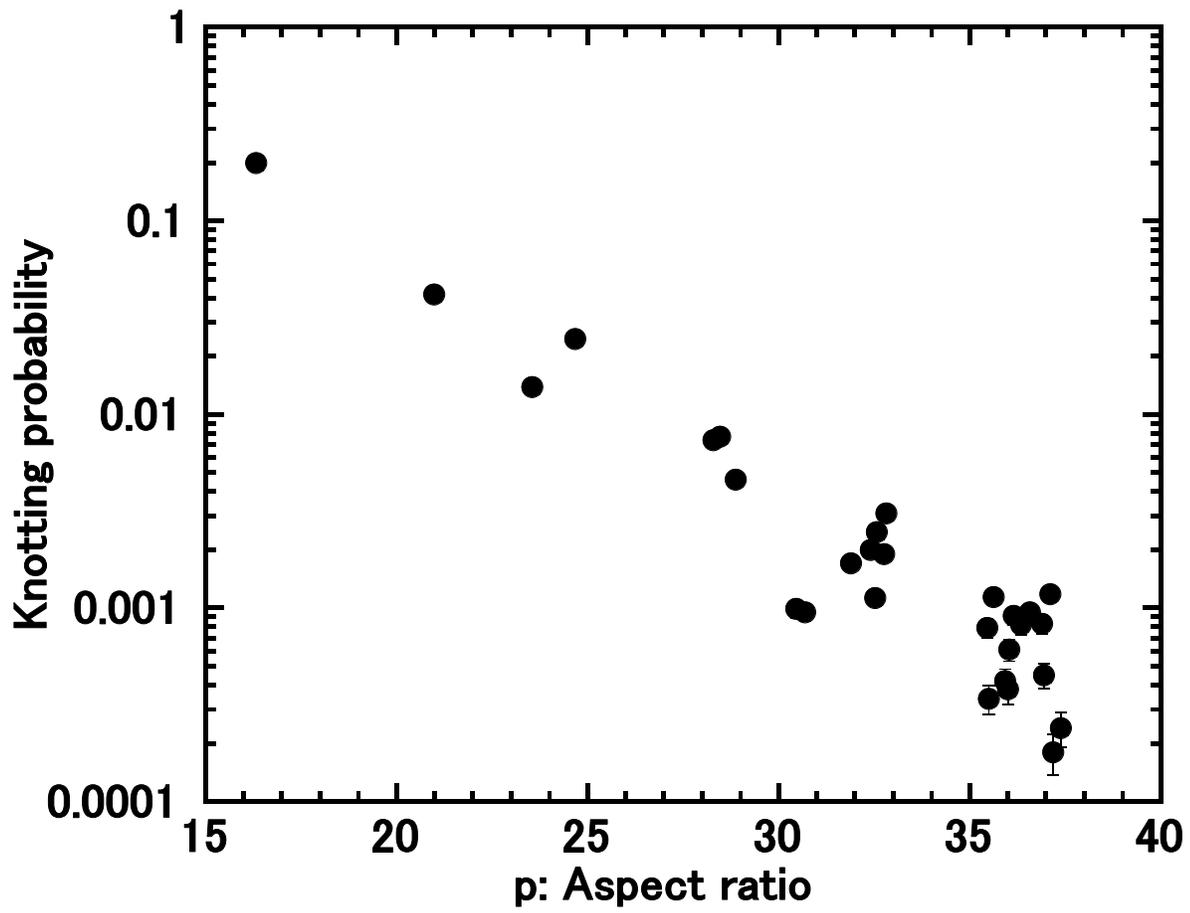
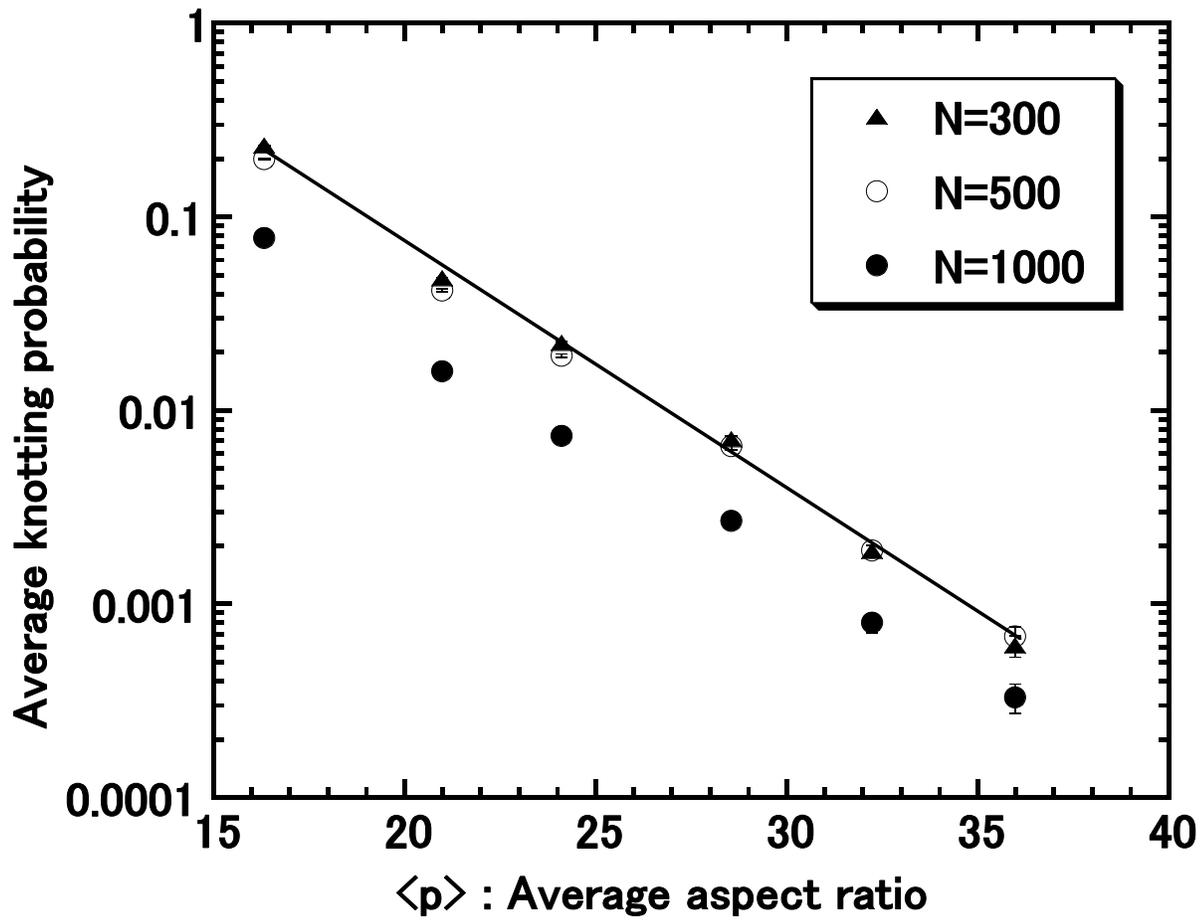
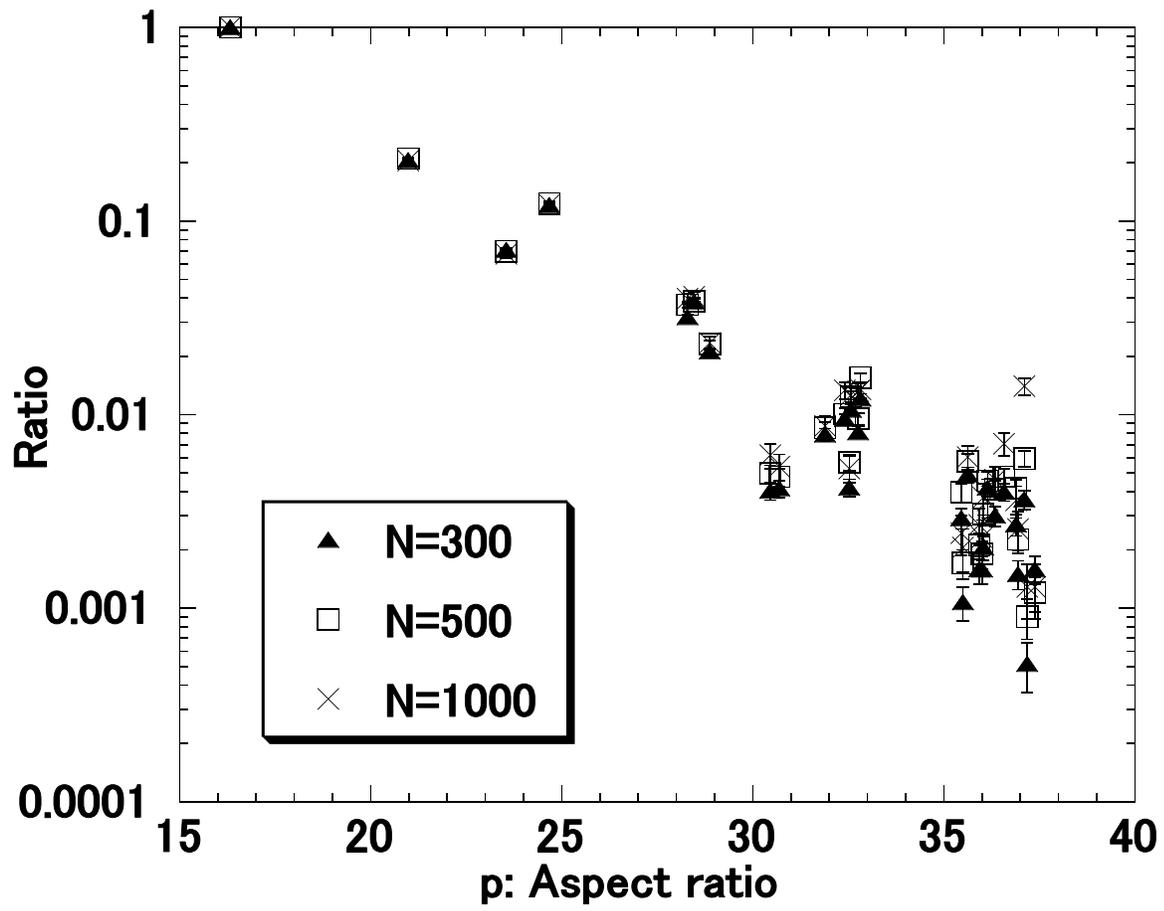
\end{document}